# CONSTRAINING SOLAR FLARE DIFFERENTIAL EMISSION MEASURES WITH EVE AND *RHESSI*

Amir Caspi[1], James M. McTiernan[2], and Harry P. Warren[3]

[1]Laboratory for Atmospheric and Space Physics, University of Colorado, Boulder, CO 80303, USA
[2]Space Sciences Laboratory University of California, Berkeley, CA 94720, USA
[3]Space Science Division, Naval Research Laboratory, Washington, DC 20375, USA

## ABSTRACT

Deriving a well-constrained differential emission measure (DEM) distribution for solar flares has historically been difficult, primarily because no single instrument is sensitive to the full range of coronal temperatures observed in flares, from $\lesssim 2$ to $\gtrsim 50$ MK. We present a new technique, combining extreme ultraviolet (EUV) spectra from the EUV Variability Experiment (EVE) onboard the *Solar Dynamics Observatory* with X-ray spectra from the *Reuven Ramaty High Energy Solar Spectroscopic Imager* (*RHESSI*), to derive, for the first time, a self-consistent, well-constrained DEM for jointly-observed solar flares. EVE is sensitive to ~2–25 MK thermal plasma emission, and *RHESSI* to $\gtrsim 10$ MK; together, the two instruments cover the full range of flare coronal plasma temperatures. We have validated the new technique on artificial test data, and apply it to two X-class flares from solar cycle 24 to determine the flare DEM and its temporal evolution; the constraints on the thermal emission derived from the EVE data also constrain the low-energy cutoff of the non-thermal electrons, a crucial parameter for flare energetics. The DEM analysis can also be used to predict the soft X-ray flux in the poorly-observed ~0.4–5 nm range, with important applications for geospace science.

## 1. INTRODUCTION

Solar flares are powerful, explosive releases of magnetic energy, heating coronal plasma to tens of megaKelvin (MK) and accelerating electrons to hundreds of MeV. The physical mechanisms behind these processes are still poorly understood (see the review by Fletcher *et al.* 2011). While it is commonly accepted that much of the hot coronal plasma results from energy deposition in the chromosphere by non-thermal, accelerated particles — "chromospheric evaporation" — X-ray observations, e.g., by *Yohkoh* or the *Reuven Ramaty High Energy Solar Spectroscopic Imager* (*RHESSI*; Lin *et al.* 2002), suggest significant *in situ* coronal heating, as well (e.g., Masuda 1994; Masuda *et al.* 1998; Caspi & Lin 2010; Longcope & Guidoni 2011; Caspi *et al.* 2014), although the specific mechanism is still debated. Another crucial problem lies in identifying the low-energy cutoff/rollover of the accelerated electron spectrum, required by energetics (see the review by Holman *et al.* 2011) but typically poorly constrained as its observable manifestation — a rollover of the non-thermal bremsstrahlung spectrum (cf. Brown 1971) — is often obscured by thermal bremsstrahlung emission that dominates the photon spectrum up to ~20–35 keV in intense flares.

These questions remain, in large part, because of the difficulty in accurately and precisely characterizing the thermal electron population in flares. While flare analyses often employ the isothermal approximation (e.g., Lin *et al.* 1981; Garcia 1994; Feldman *et al.* 1996; Holman *et al.* 2003), decades of extreme ultraviolet (EUV) and X-ray observations with widely varying temperature sensitivities have shown that flare plasma exhibits a distribution of emission measure with temperature (the "differential emission measure," or DEM). Nonetheless, there have been only a few previous studies of solar flare DEMs (e.g., Dere & Cook 1979; McTiernan *et al.* 1999; Chifor *et al.* 2007; see also Trottet *et al.* 2011, who derived a four-component discretized model for one flare); stellar DEMs have been analyzed more extensively (see, e.g., the reviews by Bowyer *et al.* 2000; Favata & Micela 2003; Güdel & Nazé 2009; and references therein).

Deriving a well-constrained flare DEM has proven difficult, largely because no single instrument's temperature response encompasses the full dynamic range of observed coronal temperatures. EUV instruments are not sensitive above ~25 MK, where the EUV-emitting ion species become depleted, and historically have suffered from poor spectral resolution (e.g., broadband filters) or coverage (e.g., narrowband filters), and/or poor temporal resolution (e.g., due to slit rastering). X-ray instruments typically become sensitivity-limited below ~10 MK, and temperatures inferred from spectral lines of hydrogen- and helium-like heavy ions (e.g., Fe, Ca, Si, etc.) often disagree significantly with those inferred from continuum emission (e.g., Phillips *et al.* 2006; Caspi & Lin 2010). A self-consistent solution has yet to be found (cf. Ryan *et al.* 2014).

The EUV Variability Experiment (EVE; Woods *et al.* 2012) onboard the *Solar Dynamics Observatory* (*SDO*; Pesnell *et al.* 2012) finally provides the means of addressing this issue comprehensively. EVE's broadband coverage, including sensitivity to all EUV-emitting ion temperatures of ~1–25 MK, combined with *RHESSI*'s sensitivity to all temperatures $\gtrsim 10$ MK, provides complete coverage of the full range of coronal plasma temperatures observed in flares with sufficient cadence to study their temporal evolution. We present a new diagnostic technique to derive a self-consistent, well-constrained DEM for jointly-observed flares, by simultaneously analyzing EVE and *RHESSI* full-Sun spectra. We apply this technique to derive the time-series DEMs for the impulsive and decay phases of two X-class flares and present the first results of the inferred thermal parameters. DEMs derived through this technique can also be used to generate synthetic soft X-ray (SXR) and EUV spectra, with important applications for the geospace community.

## 2. METHOD DETAILS

The EVE suite comprises multiple instruments, including broadband photometers and high-resolution spectrometers that together measure solar EUV/SXR emission from ~0.1 to ~105 nm (Woods *et al.* 2012). We focus on the Multiple EUV Grating Spectrograph A (MEGS-A), a grazing-incidence dual-slit spectrograph measuring the spatially integrated solar spectral irradiance from ~5 to ~37 nm, with ~0.1 nm FWHM resolution and 10 s ca-dence (Hock *et al.* 2012). The MEGS-A wavelength range includes numerous spectral lines from various ion species with formation temperatures from ~1 to ~25 MK. The 9–15 nm range is particularly rich in high-temperature lines, including most of the strongest Fe XVIII–XXIII lines (peak formation temperatures of ~6–15 MK; cf. Mazzotta *et al.* 1998); MEGS-A also observes Fe XXIV (~20 MK) at 19.204 and 25.510 nm, and Fe XV (~2 MK) and XVI (~3 MK) at 28.416 and 33.541 nm, respectively, among many other lines. This wealth of temperature diagnostics enables the most precise determination of the flare DEM to date (Warren *et al.* 2013), especially from EUV observations, although they remain poorly constrained for temperatures ≳25 MK, where EVE has little or no temperature sensitivity (Figure 1). This can be somewhat mitigated by adding observations from the X-ray Sensor (XRS) on the *Geostationary Operational Environmental Satellite* (*GOES*), but the two-channel broadband XRS measurements — from which temperatures can also be inferred (e.g., White *et al.* 2005) — add minimal spectral information and extend the DEM validity only up to ~30 MK (Warren *et al.* 2013).

Significantly stronger constraints can be obtained using *RHESSI* (Lin *et al.* 2002), which measures solar hard X-rays (HXRs) below ~0.4 nm with ≲0.1 nm FWHM resolution (quasi-constant ~1 keV when expressed in photon energy, variable $\sim\lambda^2/hc$ expressed in wavelength; Smith *et al.* 2002). *RHESSI* provides the most precise measurements of thermal X-ray continuum emission from plasmas with temperatures ≳10 MK, and is especially sensitive to the hottest part of the temperature distribution, up to ≳50 MK (e.g., Caspi *et al.* 2014). While *RHESSI* data can be used for DEM studies in isolation, *RHESSI*'s "temperature resolution" is broad because of the inherent "smoothing" of the bremsstrahlung emission mechanism; it is therefore difficult to recover a high-temperature-resolution DEM even from the high-temporal- and -spectral-resolution continuum photon spectra, and the data often cannot sufficiently distinguish between a dual-isothermal model (delta-function DEM) or a continuous DEM (e.g., Caspi & Lin 2010; McTiernan & Caspi 2014). The inferred DEM is also not well-constrained below ~10–15 MK, where the HXR yield is low and observations are typically limited by sensitivity or dynamic range (emission from the hotter parts of the DEM dominate the *RHESSI* spectra). EUV observations are required to provide these constraints.

For flares jointly observed by EVE and *RHESSI*, we have developed a technique to simultaneously analyze the full-Sun spectra from both instruments to derive a self-consistent DEM constrained over the entire range of coronal temperatures observed in flares, from ~2 MK to ≳50 MK. The lack of wavelength overlap and inherently different measurements make combining the two instruments' data non-trivial; EVE data is best analyzed versus wavelength, and *RHESSI* versus photon energy. Because the mapping from physical parameters (e.g., temperature) to observables (e.g., irradiance) is not one-to-one — a photon of a given wavelength/energy could be emitted from many different temperatures or even from non-thermal electrons — particularly for resolution-blended lines in EVE or continuum emission in *RHESSI*, direct inversion of the combined data is ill-posed and prone to large uncertainties (cf. Craig & Brown 1976), especially in regions of poor signal-to-noise (e.g., between spectral lines in EVE, or at the longest wavelengths of *RHESSI*). Since the two instruments overlap in temperature sensitivity, however, we can overcome these issues through forward modeling — adopting a physical model, deriving observables, and optimizing the model parameters to maximize agreement — and we have developed a bridge to allow simultaneous analysis of both EVE and *RHESSI* data using the methods best suited to each instrument. Figure 1 shows that the instrument inter-calibration — viz. the spectrum observed by one instrument vs. that predicted by the DEM derived independently from the other — is adequate.

Our technique combines and adapts the methods of Caspi & Lin (2010), Warren *et al.* (2013), and McTiernan & Caspi (2014). A chosen flare is subdivided into analysis intervals of ~60–120 s, providing sufficient signal-to-noise while allowing study of dynamic processes. For each interval, the 10 s cadence EVE MEGS-A Level 2 (version 4) calibrated spectra, with default ~0.02 nm binning, are temporally averaged and the lines are isolated by subtracting the underlying continuum (whether solar or instrumental); the pre-flare contribution is also subtracted. Uncertainties for each bin are derived from the standard deviation over the interval. *RHESSI* spectra are accumulated over the same interval with 1/3 keV (instrumental channel width) energy binning, and the nighttime background is subtracted; uncertainties are derived from photon counting statistics. The *RHESSI* count spectra are then pre-fit with a photon model consisting of two isothermal continua, a broken power-law non-thermal continuum, and two Gaussian features representing the Fe and Fe-Ni line complexes (cf. Phillips 2004), convolved with the instrument response, including the modifications from Caspi (2010). The model values obtained from this pre-fit provide starting guesses for the line fluxes and non-thermal parameters in the next step, below.

Our model DEM comprises multiple Gaussian components, equally spaced in log $T_e$ between 6.2 (~1.6 MK) and 7.8 (~63 MK), with fixed positions and widths but variable magnitudes; we have found that 10 Gaussians with FWHMs of ~0.19 in log $T_e$ sufficiently cover parameter space without overspecifying it. Each Gaussian is initialized to a random, uniformly distributed magnitude. Using the CHIANTI atomic database (v7.1; Dere *et al.* 1997; Landi *et al.* 2013), we compute high-resolution model spectra in the EUV (lines only) and X-ray (lines and continuum) ranges from the model DEM, then downsample to the respective instrument binning. To facilitate rapid calculations, the spectra are interpolated from a grid of CHIANTI emissivities pre-computed as functions of temperature and wavelength/energy. The model EVE spectra are convolved with its nominal ~0.1 nm FWHM spectral resolution, while the model *RHESSI* spectra are convolved with its full instrument response and the pre-fit Fe/Fe-Ni and non-thermal components are added. We then use the Levenberg-Marquardt algorithm, as implemented in the IDL SolarSoft[1] routine MPFIT (Markwardt 2009), to vary all the model parameters (Gaussian DEM magnitudes, Fe/Fe-Ni line fluxes, non-thermal power-law normalization, break energy, low-energy cutoff, and spectral indices) to iteratively minimize the deviations between the observed and model-computed spectra *of both instruments simultaneously*, as measured by the total reduced $\chi^2$ (the quadrature sum of the reduced $\chi^2$ for each instrument individually). For *RHESSI*, we use all statistically significant data from ~0.01 nm to ~0.25 nm. For EVE, we do not include the entire MEGS-A spectrum in the fit (see Figure 2), but rather focus on the line-rich 9–15 nm region, plus narrower windows around 19.2 nm, 25.5 nm, 28.4 nm, and 33.5 nm, fully covering the temperature range of ~2–25 MK. Because full-Sun irradiances of lines formed at lower temperatures (≲2 MK, e.g., Fe IX 17.1) are typically only modestly increased from the flare and, often, are *decreased* from eruption-associated coronal dimming (e.g., Mason *et al.* 2014), we do not currently analyze these cooler lines. Spatially-resolved studies (e.g., Inglis & Christe 2014) could help to better address this in the future.

---

[1] http://www.lmsal.com/solarsoft/

Because forward modeling does not guarantee uniqueness — different DEMs could potentially yield the same observed spectra, within uncertainties — we fit each interval 100 times in a Monte Carlo process. By using different model initial conditions — a new random DEM, and perturbed Fe/Fe-Ni line fluxes and non-thermal parameters — and by perturbing the observed spectra by their uncertainties for each trial, we effectively sample parameter space to test for multiple solutions (local $\chi^2$ minima) and derive uncertainties for each one; 100 trials provides a sufficient constraint on stability.

Figure 2 shows the validation and verification of our algorithm on artificial test input. Synthetic "observed" EVE and *RHESSI* spectra are generated from a contrived input DEM, with added Poisson noise to simulate real measurements. (Chromospheric lines, e.g., He II 30.4 nm, are outside the temperature range of the input DEM and are not synthesized in the EVE "observations." The *RHESSI* "observations" exclude the non-DEM model components, i.e., the Fe/Fe-Ni lines and non-thermal power-law.) The algorithm recovers the input DEM with excellent precision and accuracy, with deviations primarily in the low-emission "valleys," as expected. Convergence is rapid, arriving within 99% of the final reduced $\chi^2$ value in only nine iterations.

## 3. RESULTS AND DISCUSSION

For a first look at real data, we chose two *GOES* X-class flares — which have been shown likely to achieve the hottest temperatures (Caspi *et al.* 2014) — jointly observed by EVE and *RHESSI*: 2011 February 15 (X2.2) and 2011 March 9 (X1.5). The impulsive and early decay phase of each flare was treated as described above: subdivided into ~60 s intervals, pre-processed, then fit with a DEM (and the additional model components for *RHESSI*).

Figure 3 shows the detailed results of one run of the model fit for two time periods, during the rise and HXR peak, of the February 15 event. Overall, the model spectra fit the observations well; there are systematics in the normalized residuals for both instruments, but the results appear reasonable. Although the $\chi^2$ is high, we feel it is acceptable given that we are fitting large regions of the spectra subject to unknown systematic uncertainties. For EVE, the residuals are largest around the brightest lines, and may be due to assuming a constant 0.1 nm FWHM resolution — the actual resolution varies slightly with wavelength — with no line broadening. Additionally, chromospheric and transition region lines (~0.01–1 MK) are not modeled but may be blended with coronal lines; the most notable example is He II 25.6 nm, clearly not well-fit in the model and partly blended with Fe XXIV 25.5 nm. Higher-order contributions from the MEGS-A grating are also possibly present in the observed spectra, but are not modeled. For *RHESSI*, the residuals appear largest around the Fe/Fe-Ni line complexes, possibly due to insensitivity of the fit to these model parameters as they affect only a few out of many hundreds of data points. Despite the adequate inter-calibration (Figure 1), as we are combining data from two unique instruments, inaccuracies in the respective instrument calibrations would contribute to the combined residuals, although this is difficult to quantify. Ionization non-equilibrium is not likely to be a significant contributor to the residuals in either instrument, as the high densities (~$10^{11-12}$ cm$^{-3}$) typically observed in X-class flares (e.g., Caspi & Lin 2010; Caspi *et al.* 2014) suggest ionization timescales for Fe of $\lesssim 10$ s ($\ll 1$ s for the hotter ions, XVII–XXV; cf. Jordan 1970), much shorter than our ~60 s integration times.

Figure 4 shows the time-series average DEM fits for both events. There is a clear progression from broader, hotter DEMs earlier (rise phase and peak) to narrower, cooler DEMs later (decay phase). Conspicuously absent, however, is any significant, distinct super-hot ($T_e \gtrsim 30$ MK; log $T_e \gtrsim 7.5$ — cf. Lin *et al.* 1981) emission at any time during either flare. (The possible super-hot component during the earliest intervals of the March 9 event is weakly constrained due to poor statistics, and the *RHESSI* HXR spectra during these intervals are equally well-fit by non-thermal power-laws.) While a bimodal DEM, as has been inferred by previous studies (e.g., Jakimiec *et al.* 1988; Longcope *et al.* 2010; Caspi & Lin 2010; Warren *et al.* 2013; McTiernan & Caspi 2014), is present during many intervals, the higher-temperature peak is typically at only ~12–20 MK, well below the super-hot temperatures that would be expected from X-class events (Caspi *et al.* 2014), although we note that these two flares are fairly low X-class. The peaks or plateaus in the joint-instrument DEM are typically at lower temperatures than the dual-isothermal approximations obtained from the *RHESSI*-only pre-fits, which is not unexpected since the continuous DEM predicts additional HXR flux from temperatures between and below the isothermally-approximated values, requiring less high-temperature emission in the DEM to explain the remaining flux observed by *RHESSI*.

Because our model includes a non-thermal component, we are able to obtain limits on the non-thermal low-energy cutoff (Holman *et al.* 2011). Importantly, the EVE data provide a *lower* bound for the cutoff energy, unobtainable from *RHESSI* alone, as pushing the cutoff too low reduces the thermal model below what is supported by the EVE spectra; this is currently under study.

## 4. SUMMARY AND CONCLUSIONS

*SDO*/EVE's unprecedented solar EUV observations, coupled with *RHESSI*'s groundbreaking X-ray measurements, provide an ideal toolkit to study the evolution of both thermal plasma and accelerated, non-thermal electrons in solar flares.

We have developed a technique to simultaneously analyze the EUV and X-ray spectra from EVE and *RHESSI*, respectively, enabling derivation of the most accurate, self-consistent flare DEMs to date covering the entire coronal temperature range observed in flares, ~2 to $\gtrsim 50$ MK. We have validated this algorithm on artificial test data, and applied it to two X-class flares from the current solar cycle. The derived time-series DEMs progress clearly from broad, hot distributions to narrower, cooler distributions, and a bimodal signature is often observed. The DEMs suggest that fitting *RHESSI* data in isolation, particularly with isothermals, may bias the results towards higher temperatures (cf. Ryan *et al.* 2014), highlighting the need for the additional constraints provided by EVE. Indeed, these DEMs show little evidence of a significant super-hot component, as would be naïvely expected from X-class flares based on the results of Caspi *et al.* (2014). However, we note that most of the *RHESSI* dual-isothermal pre-fits also yielded no super-hot component, the only exceptions being the earliest one or two intervals of each event, which also had the broadest and highest-temperature DEMs; these two flares may simply be cooler events. Further, despite their X-class intensity, both flares were highly impulsive, short-duration events with little or no HXR emission $\gtrsim 100$ keV; such flares may experience different heating than the eruptive, long-duration events more typically seen with these intensities (such as 10 of the 12 X-class flares in Caspi *et al.* 2014). The 2011 August 9 X6.9 flare, a longer-duration eruptive event, is currently under study and will help address this question.

One essential application of this DEM analysis will be to constrain the non-thermal electron low-energy cutoff. Comparing the thermal and non-thermal electron populations, and their evolution, would help to test different theories of plasma heating or to evaluate the partition between different heating mechanisms (e.g., Liu *et* *al.* 2013; Oreshina & Oreshina 2013).

Our technique also has important geospace applications, in helping determine the solar SXR loading on Earth's upper atmosphere. The wavelength range of ~0.4–5 nm is thought to contain the bulk of flare-radiated energy (Rodgers *et al.* 2006) and is the primary driver of ionospheric dynamics. Very few spectrally resolved observations exist in this range, and the detailed dynamics depend critically on the altitude of energy deposition, strongly determined by the (unknown) spectral energy distribution (e.g., Sojka *et al.* 2013). Our physically modeled DEMs allow generation of synthetic X-ray spectra in this unobserved passband, filling a crucial gap in the needed input for modeling of upper atmospheric behavior.

This work was supported by NASA grant NNX12AH48G. A. Caspi and H. Warren were also supported by NASA contract NAS5-02140, and J. McTiernan by NASA contract NAS5-98033.

## REFERENCES

Bowyer, S., Drake, J. J., & Vennes, S. 2000, *ARA&A*, **38**, 231
Brown, J. C. 1971, *SoPh*, **18**, 489
Caspi, A. 2010, *PhD thesis*, Univ. California, Berkeley (arXiv: 1105.1889)
Caspi, A., Krucker, S., & Lin, R. P. 2014, *ApJ*, **781**, 43
Caspi, A., & Lin, R. P. 2010, *ApJL*, **725**, L161
Chifor, C., Del Zanna, G., Mason, H. E., *et al.* 2007, *A&A*, **462**, 323
Craig, I. J. D., & Brown, J. C. 1976, *A&A*, **49**, 239
Dere, K. P., & Cook, J. W. 1979, *ApJ*, **229**, 772
Dere, K. P., Landi, E., Mason, H. E., Monsignori Fossi, B. C., & Young, P. R. 1997, *A&A*, **125**, 149
Favata, F., & Micela, G. 2003, *SSRv*, **108**, 577
Feldman, U., Doschek, G. A., Behring, W. E., & Phillips, K. J. H. 1996, *ApJ*, **460**, 1034
Fletcher, L., Dennis, B. R., Hudson, H. S., *et al.* 2011, *SSRv*, **159**, 19
Garcia, H. A. 1994, *SoPh*, **154**, 275
Güdel, M., & Nazé, Y. 2009, *A&ARv*, **17**, 309
Hock, R. A., Chamberlin, P. C., Woods, T. N., *et al.* 2012, *SoPh*, **275**, 145
Holman, G. D., Aschwanden, M. J., Aurass, H., *et al.* 2011, *SSRv*, **159**, 107
Holman, G. D., Sui, L., Schwartz, R. A., & Emslie, A. G. 2003, *ApJL*, **595**, L97
Inglis, A. R., & Christe, S. D. 2014, *ApJ*, in press (arXiv: 1405.5262)
Jakimiec, J., Pres, P., Fludra, A., Bentley, R. D., & Lemen, J. R. 1988, *AdSpR*, **8**, 231
Jordan, C. 1970, *MNRAS* **148**, 17
Landi, E., Young, P. R., Dere, K. P., Del Zanna, G., & Mason, H. E. 2013, *ApJ*, **763**, 86
Lin, R. P., Dennis, B. R., Hurford, G. J., *et al.* 2002, *SoPh*, **210**, 3
Lin, R. P., Schwartz, R. A., Pelling, R. M., & Hurley, K. C. 1981, *ApJL*, **251**, L109
Liu, W.-J., Qiu, J., Longcope, D. W., & Caspi, A. 2013, *ApJ*, **770**, 111
Longcope, D. W., Des Jardins, A. C., Carranza-Fulmer, T., & Qiu, J. 2010, *SoPh*, **267**, 107
Longcope, D. W., & Guidoni, S. E. 2011, *ApJ*, **740**, 73
Markwardt, C. B. 2009, in *ASP Conf. Ser. 411, Astronomical Data Analysis Software and Systems XVIII*, ed. D. A. Bohlender, D. Durand, & P. Dowler (San Francisco, CA: ASP), 251
Mason, J. P., Woods, T. N., Caspi, A., Thompson, B. J., & Hock, R. A. 2014, *ApJ*, in press (arXiv: 1404.1364)
Masuda, S. 1994, *PhD thesis*, Univ. Tokyo
Masuda, S., Kosugi, T., Sakao, T., & Sato, J. 1998 in *Observational Plasma Astrophysics: Five Years of Yohkoh and Beyond*, ed. T. Watanabe, T. Kosugi, & A. C. Sterling (Boston, MA: Kluwer), 259
Mazzotta, P., Mazzitelli, G., Colafrancesco, S., & Vittorio, N. 1998, *A&A*, **133**, 403
McTiernan, J. M. & Caspi, A. 2014, *ApJ*, in preparation
McTiernan, J. M., Fisher, G. H., & Li, P. 1999, *ApJ*, **514**, 472
Oreshina, A. V., & Oreshina, I. V. 2013, *A&A*, **558**, A16
Pesnell, W. D., Thompson, B. J., & Chamberlin, P. C. 2012, *SoPh*, **275**, 3
Phillips, K. J. H. 2004, *ApJ*, **605**, 921
Phillips, K. J. H., Chifor, C., & Dennis, B. R. 2006, *ApJ*, **647**, 1480
Rodgers, E. M., Bailey, S. M., Warren, H. P., Woods, T. N., & Eparvier, F. G. 2006, *JGRA*, **111**, 10
Ryan, D. F., O'Flannagain, A. M., Aschwanden, M. J., & Gallagher, P. T. 2014, *SoPh*, **289**, 2547
Sojka, J. J., Jensen, J., David, M., *et al.* 2013, *JGRA*, **118**, 5379
Smith, D. M., Lin, R. P., Turin, P., *et al.* 2002, *SoPh*, **210**, 33
Trottet, G., Raulin, J.-P., Giménez de Castro, G., *et al.* 2011, *SoPh*, **273**, 339
Warren, H. P., Mariska, J. T., & Doschek, G. A. 2013, *ApJ*, **770**, 116
White, S. M., Thomas, R. J., & Schwartz, R. A. 2005, *SoPh*, **227**, 231
Woods, T. N., Eparvier, F. G., Hock, R., *et al.* 2012, *SoPh*, **275**, 115

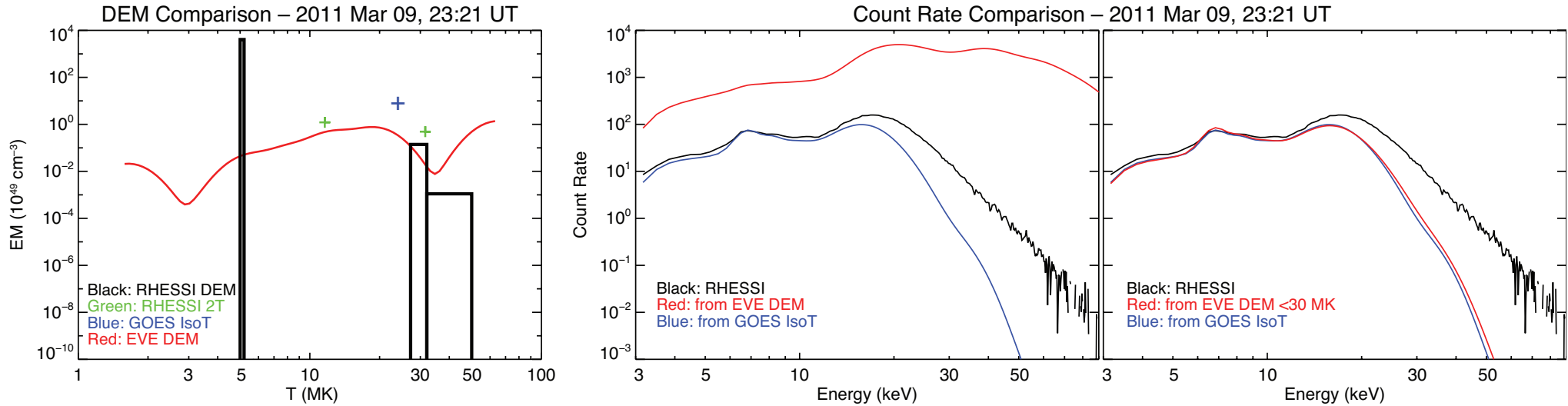

**Figure 1.** [*left*] DEMs derived from *RHESSI* (*black*) and EVE (*red*) data in isolation, with *RHESSI* dual-isothermal pre-fit (*green plusses*) and *GOES* XRS isothermal (*blue plus*) for comparison. The *RHESSI* DEM has limited resolution and poor constraints ≲10 MK; the EVE DEM is better resolved but poorly constrained ≳25 MK. [*center*] Observed *RHESSI* count rate spectrum vs. photon energy (*black*) and synthetic models derived from the EVE DEM (*red*) and *GOES* isothermal (*blue*); the high-temperature excess in the EVE-derived model overestimates observations by ≳10×. The *GOES*-derived model agrees well at low energies, but indicates need for a higher-temperature component, consistent with the dual-isothermal fit. [*right*] Arbitrarily truncating the EVE DEM at ~30 MK yield good model agreement with *RHESSI* at lower energies, consistent with the *GOES*-derived model, and suggests that the instrument inter-calibration is adequate.

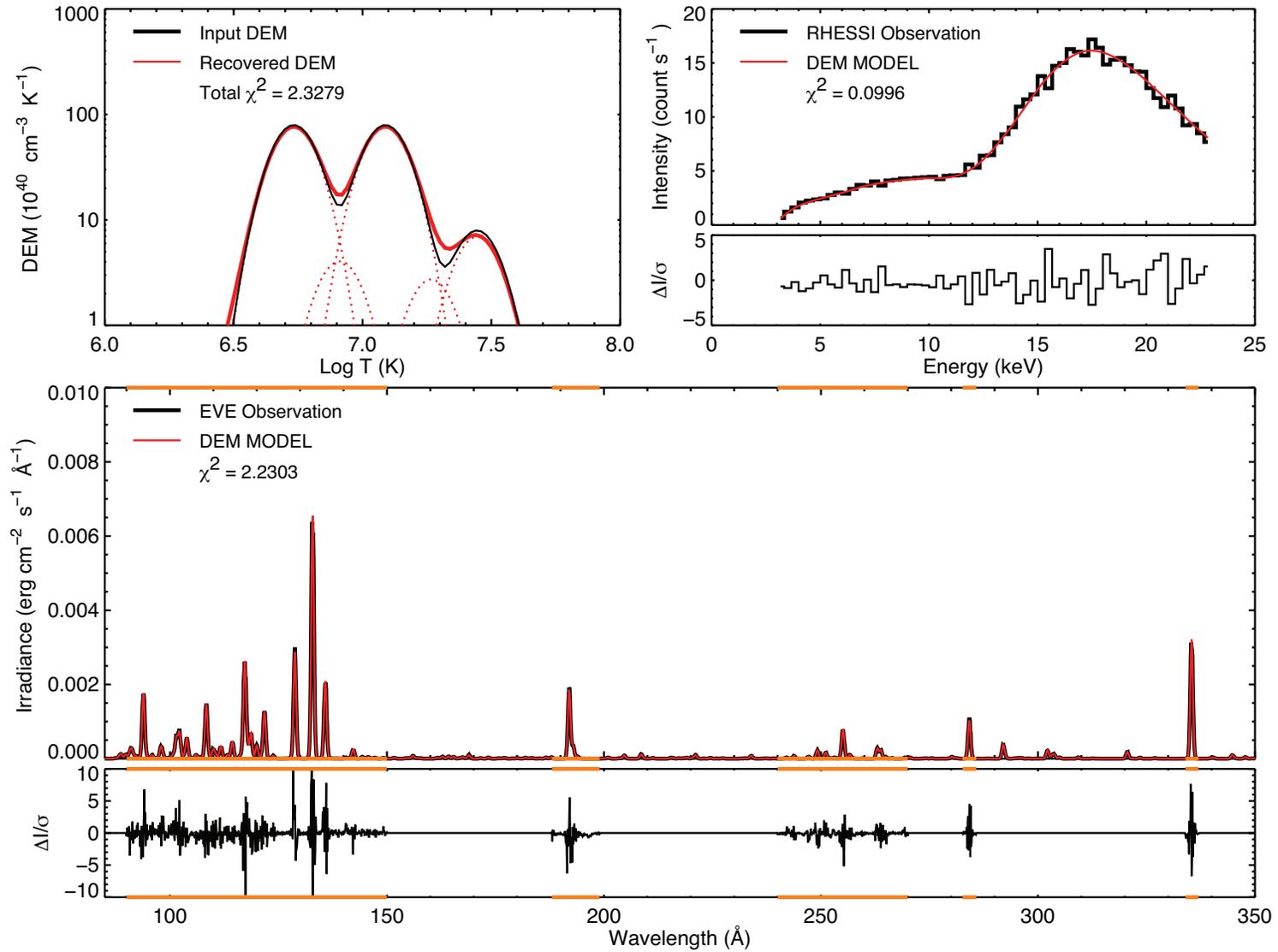

**Figure 2.** Algorithm validation on artificial test input. Synthetic EVE (*bottom*) and *RHESSI* (*top right*) spectra are generated from the input DEM (*top left*), with added Poisson noise; the DEM is recovered by forward modeling. *Orange* highlights indicate the EVE wavelengths included in the fit; chromospheric lines are not synthesized. The recovered DEM (*red*), with constituent components shown (*dotted*), matches the input (*black*) well.

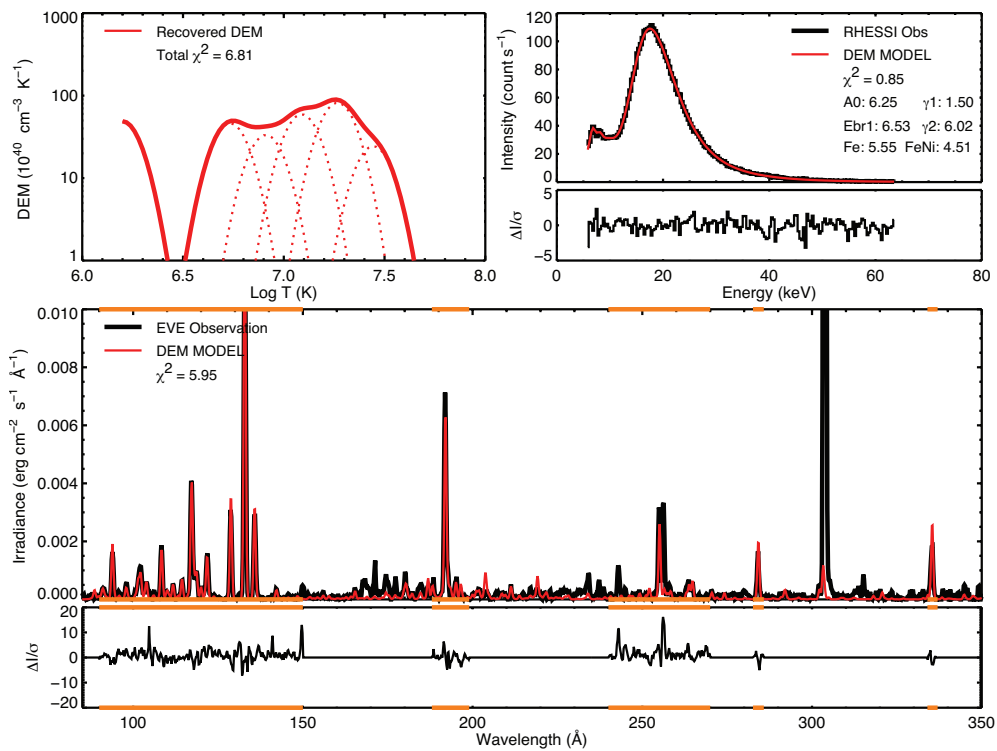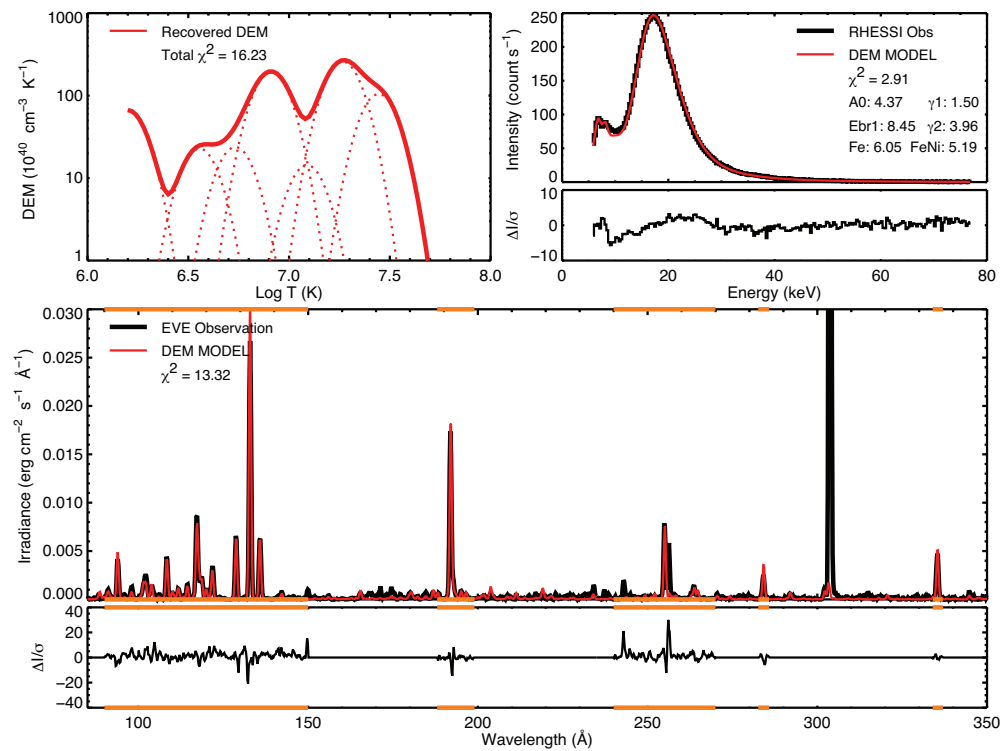

**Figure 3.** DEM fits from two intervals of the 2011 February 15 X2.2 event, during the rise (~01:51 UTC; *left*) and HXR peak (~01:53 UTC; *right*). Systematics exist in the normalized residuals for both instruments, but the fits are reasonable, with acceptable $\chi^2$ considering that entire spectral regions are being fit and accounting for modeling limitations.

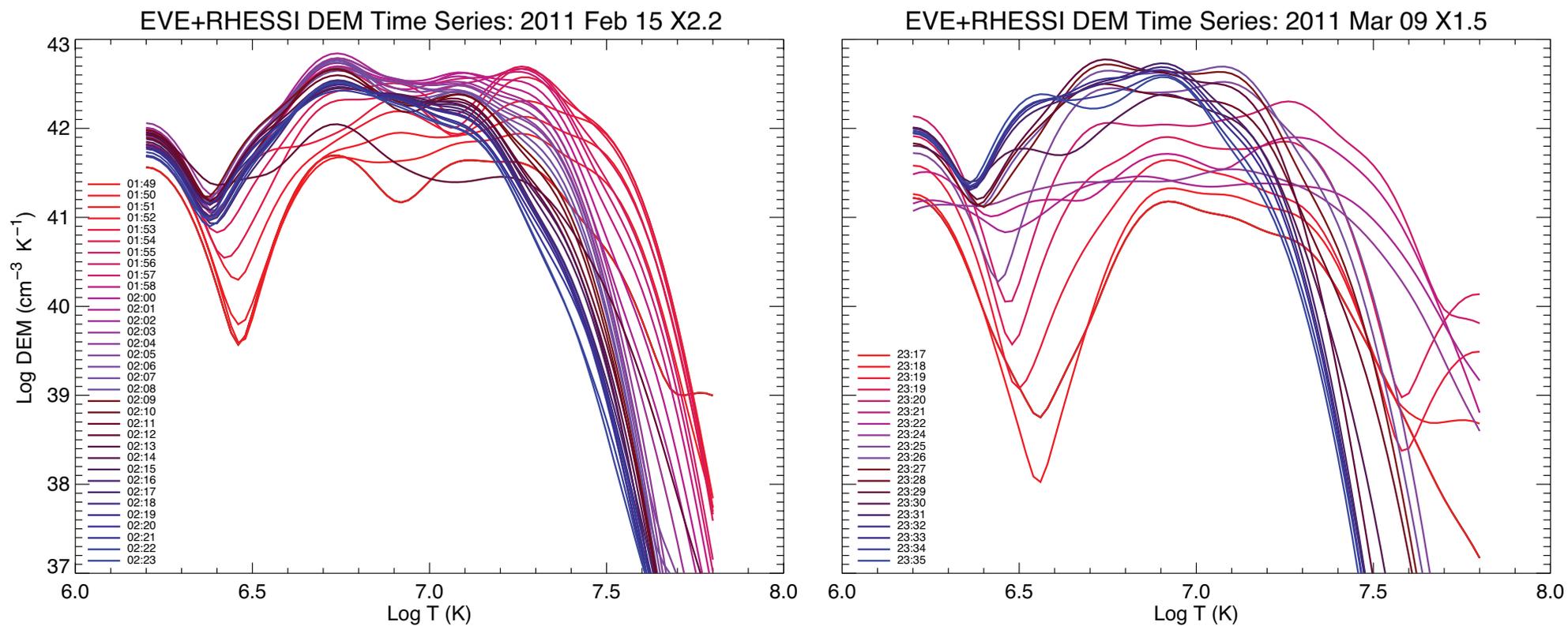

**Figure 4.** Time-series DEM fits from the 2011 February 15 X2.2 (*left*) and 2011 March 9 X1.5 (*right*) flares. Each curve represents the Monte Carlo average DEM for that interval; uncertainties are omitted for clarity. A clear progression from hotter, broader DEMs to cooler, narrowed DEMs is evident, with bimodal distributions in many intervals but with no obviously distinct, significant super-hot (log $T_e \gtrsim 7.5$) component at any time (the beginning of March 9 is only weakly constrained due to low *RHESSI* count statistics).